\documentstyle[preprint,aps]{revtex}
\begin{document}

\title{
Generalized Lotka-Volterra (GLV) Models
 }
\author{
Sorin Solomon
\footnote{
To appear in Econophysics 97, 
(Kluver 1998), eds. Imre Kondor and Janos Kertes
}
}
\address{
Racah Institute of Physics,
The Hebrew University,
Jerusalem 91904,
Israel
}
\maketitle

\begin{abstract}

 The Generalized Lotka-Volterra (GLV) model:

    $$  {w_i(t+1)} = {\lambda w_i(t) + a {\bar w}(t) - c {\bar w}(t) w_i(t)}
\ \ \ , \ \ i=1, ......, N  $$

 provides a general method to simulate, analyze and understand a wide
 class of phenomena that are characterized by power-law probability
 distributions:

    $$  {P(w) dw} \sim {w^{-1 -\alpha} dw}  \quad (\alpha \geq 1)  $$

 and truncated Levy flights fluctuations $L_{\alpha}({\bar w})$.

\end{abstract}

\section{Introduction}

 Many natural and man-made phenomena are known to involve power-law
 probability distributions (e.g. Pareto 1897; Zipf 1949; Mandelbrot
 1961, 1951, 1963; Atkinson and Harrison 1978; Bak et al. 1997;
Bouchaud et al. 97;
Cahalan and Joseph 1989;
 Mantegna and Stanley 1994, 1995, 1996, 1997; 
Stanley et al. 1995; Sornette et.al. 1997;
 Zanette and Manrubia 1997, Zhang et al 1997).

 Power-laws are common in systems composed of units that have no
 characteristic size, and in systems made of auto-catalytic elements
 (Yule 1924; Champernowne 1953; Simon and Bonini 1958;
Ijiri and Simon 1977, Anderson 1995).
 Moreover it has been shown that in systems which are not separable
 into energetically independent parts, the power laws take naturally the
 place of the usual exponential (Boltzmann) distribution (Tsallis 1988).

 It has been shown both theoretically (Solomon 1998; Solomon and Levy 1996)
 and using numerical simulations (Biham et al. 1998) that the Generalized
 Lotka-Volterra (GLV) model produces power-law distributions,
 but doesn't suffer from the problems and limitations of previous
 similar models.  Since the GLV requires only a few, and not too
 restrictive pre-conditions, it may be expected to be widely applicable.

 The GLV involves three scalar parameters, and a probability distribution,
 each having clear roles in the model's interpretation.  Certain model
 properties are "universal" i.e. independent of some of the model's
 parameters.  This conceptual simplicity makes the GLV easily adaptable
 to different systems, and makes it possible to use it as a descriptive,
 understanding and comparison tool.

 The features listed above make the GLV a candidate for a general
 method to simulate, analyze and understand a wide class of phenomena
 which are characterized by power-law probability distributions
and multiscale fluctuations.

 The aim of this paper is to provide a practical introduction to
 GLV modeling.  We describe the basic theory, how to construct
 a numerical simulation of the GLV and applications to practical
 interesting systems.

 For didactic reasons we will present a few previous and simpler
 models before introducing the GLV. This will nicely partition
 the topics involved into more manageable parts.  The
 models described in this paper are:

 \begin{itemize}

 \item{}  Single-agent random multiplicative process without barrier eq. (\ref{lab1}).
\item{}  Single-agent multiplicative process with fixed lower barrier eq. (\ref{lab11}).
\item{}  Multiple agent process with barrier coupled to the mean eq. (\ref{lab28}-\ref{lab31}).
\item{} Single-agent linear (stochastic multiplicative/additive) process eq. (\ref{kesten})
   \item{}  Generalized Lotka-Volterra (autocatalytic competing) agents eq. (\ref{lastglv}).
 \end{itemize}

\section{Some basic concepts}

 Microscopic representation or agent-oriented simulation is a relatively new way of
 generating and expressing knowledge about complex systems
 (see Solomon 1995 for a physicist oriented review and Kim and  Markowitz 1989  for an early
pioneering work in finance and ZhangWB 1991 for a synergetically inspired view).

 One idea this document tries to convey is the basic difference
 between standard methods of explanation, which are based on a global
 parameterization of a macroscopic dynamics and the ``microscopic
 representation'' or ``agent-oriented'' simulation method of explanation.

 To illustrate the older methods, think of the high-school textbook
 problem of computing the trajectory of a stone under the influence
 of gravitation.  The stone is treated as a point mass (that is
 the global parameterization), and we use Newton's Laws (that's
 the macroscopic dynamics).

 This kind of explanation is not always useful.
For instance in order to study the cracking  and crumbling of  the stone
under pressure, one would have to consider its structure made of clusters
of  (of clusters ...) of smaller stone parts.   In many physical,
 economic etc situations, we find it is most natural to view the
 system under study as a collection of ``microscopic'' similar units
 (or ``agents''), which interact among themselves in some well-defined
 way and (co-)evolve in time.

 Experience have shown that even a simple dynamics applied to a
 system consisting of many similar interacting agents may produce
 interesting macroscopic effects.  The properties which arise
 from the collective behavior of many similar elements are
 called emergent properties.

 Some examples of systems with emergent properties are:

  Animal (and human) populations (cities/countries) are
          composed of individuals which are constantly born (and die),
          and compete with other members of their species
          (for food, mates etc).

          The classical (scalar) Lotka-Volterra system:

 \begin{equation}
w (t+1) = (1 + {\rm birth} - {\rm death}) * w (t) - {\rm competition} * w (t)^2
 \label{LV}
 \end{equation}
 was    invented specifically in order to model
the time evolution of such populations $w(t)$.
We will show that its straightforward multi-agent generalization eq. (2)
which treats the population as a collection of sub-populations $w_i$,
presents emergent properties very different from the eq. (1) which treats
the entire population as a single variable (similarly to the
stone/point mass example above).

  The stock market is parametrized macroscopically by the
 index $w (t)$ which is proportional to the
 capitalization (total worth of the money invested) in the traded
 equities. For a fixed amount of shares, this is a measure of the
 price of the shares traded in the market.
 Models similar to the stone/point-mass/Newton law example above were
 introduced in the past (Baillie and  Bollerslev 1990)
 and lead to models which treat the index/share price $w (t+1)$
as a single entity
 with time evolution governed by stochastic differential equations.

 The microscopic representation method models the market index $\bar w (t)$
 as an collective quantity emerging from the interactions of a
macroscopic number of traders.

 More precisely, the "microscopic representation" of the stock market
 is composed of many investors $i=1 , ...., N$
          each having a certain personal wealth $w_i (t)$.  The investors
          buy and sell stocks, using more or less complicated
          strategies designed to maximize their wealth.
The resulting index is then proportional to the total sum of the
individual investors wealth.

          Another ("orthogonal") way to characterize the financial markets
          is to consider as elementary degrees of freedom
          the individual stocks (i.e. the capitalization $w_i (t)$ of the
          individual companies $i$ traded in the market).

          We are going to use the Generalized Lotka-Volterra (GLV)
          equation system
\begin{equation}
      {w_i(t+1)} = {\lambda(t) w_i(t) + a(t) {\bar w(t)} - c(t) {\bar w(t)}  w_i(t)}
\ \ \ , \ \ i=1, ....., N
\label{GLV}
 \end{equation}
to construct a very simple model
          of the stock market.  In spite of being very simple,
          the model yields (and in doing so - explains) the Pareto
          distribution of wealth, an economical fact left for
          a long time (100 years!) without satisfactory explanation.
In the case of the previous example of populations/cities/countries sizes,
the system eq. (\ref{GLV}) results from the
assumption that the $w_i$ individuals belonging to each city/country
$i$ interact similarly (in the stochastic sense)
 to all the individual members of the other cities/countries.
 More precisely, eq. (\ref{GLV}) would result from
 the following assumptions:

  - for each of the $w_i$ citizen of a city,
 there is a $\lambda$ probability that he will
 attract (or give birth to) a new citizen.
 One may include as a negative contribution to $\lambda$
 the probability for the citizen to die or leave the town
 without preferred destination.

  - each of the $w_i$ citizens of a city has a probability
  $c/N$ to be attracted by one of the $N \bar w$ citizens of the
  existing cities and leave the town $i$.

  - the citizens which left own town but are not bound to a particular town,
  have the probability $a$ to end up in any of the $i$ towns.

 The analysis below shows that if the last term is negligible,
 the exponent $\alpha$ is likely to be $1$.

 Systems made of auto-catalytic (terms $\lambda $ and $a$)
competing (term $c$) elements are wide-spread in nature.
Another example, very different from the previous:
clouds are collections of water droplets, which grow
          by taking in water vapor from the surrounding air.
          Agent-oriented models may explain the observed
          distribution of cloud sizes and shapes without going
          into all the very complicated physical details.

 To make things more concrete, we will often use in the following
 discussion terminology borrowed from the GLV model of the stock market.
 However, our results and insights will be quite generic and applicable
 to other systems.

 For example, we will often use the term `traders' $i$ instead of
 microscopic elements and talk about the `wealth' $w_i$ of each trader
 and the "average wealth" $\bar w (t)$:
\begin{equation}
      {\bar w} (t) = {1\over N} \sum_i w_i(t)
\label{total}
 \end{equation}
 All these names should be suitably renamed
 when used in other contexts.

\section{Probability Distributions as Predictive Output of Modeling}

 Modeling the stochastic evolution in time  $t= 1, .....,T$
of a system of type eq. (\ref{LV})
 produces a lot of raw data.  It is not very useful
 to produce more and more sequences of numbers
$$ w (t) ; \  \ t = 1, ...., T $$
without a way to analyze this data.

 A natural way is to collect a lot of such sequences and analyze them
 statistically. A way to look at it is to imagine a large set of (uncoupled)
 traders $w_i, \ \ i= 1, ..., N$, subject to the same stochastic dynamics, and compute their
 distribution.
More precisely, assuming each data sequence $i$
produced by eq. (\ref{LV}) describes the evolution of one trader $i$
$$ w_i (t) ; \ \  t = 1, ...., T $$
we may compute  over a large set of traders $i= 1, ...., N$
the number of traders $N_t (w)$  with wealth
$w_i (t)$ in the interval $( w, w+ dw)$.
In the limit of infinite $N$ this defines
the individual wealth probability distribution
\begin{equation}
      P_t (w ) dw = N_t (w) /N
\label{distr}
 \end{equation}

This way of looking at the problem is conductive to a model
like  eq. (\ref{GLV})
in which the various traders {\bf do} interact e.g. through terms
 including their average $\bar w$.  As we will see, this is a key
 ingredient in solving some of the problems with the one-agent models
of the type eq. (\ref{LV})
 and of course a way to make the models more realistic.

\section{The general shape of the probability distributions}

 It may be a little surprising, but in certain conditions, even for
 non-stationary interacting non-linear dynamical systems of the type eq. (\ref{GLV}),
we may predict (without performing
 any computer runs) how the wealth distributions $P_t (w)$ produced by our
 models will look like.

 Before going into a more detailed analysis we list here some obvious
 properties which the probability distribution has to obey.

 \begin{itemize}
 \item{}  The distribution vanishes for negative values.
$$ P (w) = 0 \ for \ w <0 $$
          This property follows from the way we choose the
          initial wealth values, and the nature of the dynamics.
          In practice this corresponds to the fact that
population sizes, capitalization of companies, cloud sizes, etc. assume only
          positive values.
 \item{}  The value zero usually has vanishing probability
          distribution $P (0) = 0$.
          This follows from the continuity of the distribution,
          and the previous characteristic.
 \item{}  The distribution has a tail which may obey a
          power-law
\begin{equation}
      P (w ) \sim w^{-1-\alpha}
\label{powel}
 \end{equation}
or not. In any case, distributions must decay to zero at infinity:
 \begin{equation}
   \lim_{w->\infty}   P (w ) = 0
\label{tail}
 \end{equation}
          otherwise the probability to have finite $w_i$
values would vanish.
 \item{}  The distribution has a maximum.
          A ``nice'' continuous and non-negative function will
          have a maximum between two zeroes.  Here we have one
          at zero, and one at infinity.
 \end{itemize}
 \medskip
 The remaining non-trivial issues are:
 \begin{itemize}
   \item{}  The behavior of the tail, in particular whether
            it decreases as a power-law eq. (\ref{powel}), log-normal,
            exponential or some other shape.
   \item{}  The location and height of the maximum
   \item{}  The behavior of the rising part
   \item{}  The relation between the tail shape and the parameters of the model.
We will see that in certain conditions, the tail behavior is quite universal
in as far as it is quantitatively independent on most of the parameters and it is preserved even
in generically non-stationary conditions.
 \end{itemize}

 In this article we will be interested mainly in the shape of the tail
 of the probability distribution $P (w)$, in particular whether it is  a power-law
 eq. (\ref {powel}) or not.

The importance of the power laws is far from purely academic:
the power laws are the bridges between the
simple microscopic elementary laws acting at the individual level and
the complex macroscopic phenomena acting at the collective level.
They insure that the dynamics does cover many dynamical
scales rather than acting only at  the smallest and/or largest scales of the system.

Moreover, power laws insure that the dynamics
of  the intermediate scales of the system is largely independent on the
microscopic details and the macroscopic external conditions constraining the
system. Consequently the macroscopic complexity is not the direct and
linear result of the microscopic details but rather a generic consequence of the
self-organization taking place in systems with many interacting parts
and inter-related feedback loops.

This means that one can hope to extract macroscopic and mesoscopic laws
which hold for large classes of microscopic laws in an universal way.

This unifying and predictive power of the power laws made them the
central objects in many branches of science starting with quantum field theory and
statistical mechanics and ending with ecology and economics.
In fact the necessity for such laws in fundamental physics lead to very
"un-natural" (t'Hooft 1979) fine-tuning procedures designed to enforce them.

One can say that the emergence of the power laws is a {\it sine qua non} condition for the
emergence of the  macroscopic world out of local microscopic elementary laws of nature.
As such, it becomes a fundamental natural law on its own.

{\bf The present paper shows that power laws emerge generically in the most simple and natural
models which were considered in the past in the modeling of chemical, biological and social
systems}.

\section{Multiplicative random dynamics and log-normal distributions}

 Power-law probability distributions eq. (\ref{powel}) with exponent $-1$ (i.e. $\alpha = 0$),
 can be obtained analytically from a multiplicative process (Redner 1990;
 Shlesinger 1982):
 \begin{equation}
    w(t+1) = \lambda(t) w(t)
 \label{lab1}
 \end{equation}
 where the random variables $\lambda(t)$ are extracted from a fixed
 probability distribution $\Pi (\lambda)$ with positive support.

 Indeed, in order to obtain the distribution of $w(t)$ in the large
 $t$ limit, one takes the logarithm on both sides of eq. (\ref{lab1}) and
 uses the notations $\mu = \ln \lambda$, $x = \ln w$.
 With these notations eq. (\ref{lab1}) becomes:
 \begin{equation}
    x(t+1) = \mu (t) +  x(t)
 \label{lab6}
 \end{equation}
 The respective probability distributions $\rho (\mu )$ and
 ${\cal P} (x)$ for $\mu$ and $x$ are related to the distributions
 for $\lambda$ and $w$ by the identities:
 \begin{equation}
   {\rho} (\ln \lambda) d (\ln \lambda ) = \Pi (\lambda) d \lambda
 \label{lab2}
 \end{equation}
 and respectively
 \begin{equation}
   {\cal P} (\ln w) d (\ln w) = P (w) d w
 \label{lab3}
 \end{equation}
 which mean:
 \begin{equation}
   {\rho} (\mu) d \mu  = ({\exp {\mu}}) \Pi ({\exp{\mu}}) d \mu
 \label{lab4}
 \end{equation}
 and respectively
 \begin{equation}
   {\cal P} (x) d x  = ({\exp {x}}) P ({\exp{x}}) d x
 \label{lab5}
 \end{equation}

The interpretation of eq. (\ref{lab6}) is that
 $x(t+1)$ is the sum of the constant $x(0) = \ln w(0)$, and $t$
 random variables $\mu (t) = \ln \lambda (t)$ extracted from the fixed
 distribution $\rho (\mu )$.  Under the general conditions of the
 Central Limit Theorem (CLT) we get for $x(t)$ at large $t$ the
 normal distribution:
 \begin{equation}
    {\cal P}_t (x) \sim
    {1 \over \sqrt {2 \pi \sigma^2_{\ln \lambda} t}}
      \exp {- {{(x - <x>)^2} \over {2 \sigma^2_{\ln \lambda} t}}}
 \label{lab7}
 \end{equation}
 where
 \begin{equation}
    <x> = t v \equiv t <\ln \lambda>
 \end{equation}
 and
 \begin{equation}
    \sigma^2_{\ln \lambda} = < (\ln \lambda)^2> - <\ln \lambda>^2
 \end{equation}


 Note that the width of the normal distribution ${\cal P}_t (x)$
(the denominator
 of the expression in the exponential in eq. (\ref{lab7})) is
 \begin{equation}
     \sigma^2_x = t \sigma^2_{\ln \lambda}
 \end{equation}
 and increases indefinitely with time. This means that the distribution  ${\cal P}_t (x)$
 becomes independent of $x$
 \begin{equation}
   {\cal P}_t (x) \sim  {1 \over \sqrt {2 \pi \sigma^2_{\ln \lambda} t}}
 \label{lab8}
 \end{equation}
 in an ever-increasing neighborhood of $<x>$ (where the exponential is close to $1$).

 Transforming eq. (\ref{lab8}) back to the $w$ variables by using
 eq. (\ref{lab3}) one gets
 \begin{equation}
    P (w) d w = {1 \over \sqrt {2 \pi \sigma^2_{\ln \lambda} t}} d (\ln w)
 \label{lab9}
 \end{equation}
 i.e.
 \begin{equation}
    P (w) dw  \sim   1/w  \  dw
 \label{lab10}
 \end{equation}

 Graphically, this means that
as the time goes to infinity,
 $P(w)$ is an ever ``flattening'' distribution approaching $w^{-1}$
 in an ever expanding neighborhood.


\section{Non-interactive
         multiplicative processes with fixed lower bound}

 In order to obtain a power-law probability distributions
 $P (w) \sim w^{-1 -\alpha}$ with exponent $\alpha > 0$ the
 multiplicative process
eq. (\ref{lab1})
 has to be modified
(Yule 1924, Champernowne 1953, Simon and Bonini 1958, Ijiri and Simon 1977)
in such a way that the variation of  $w(t)$  under
eq. (\ref{lab1}) will be constrained by a  lower bound (or barrier) $w_{min}$
 \begin{equation}
     w (t) > w_{min} .
 \label{lab11}
 \end{equation}

In terms of $x (t) = \ln w (t)$ , eq. (8) becomes supplemented by the lower bound:
 \begin{equation}
     x (t) > x_{min} \equiv \ln (w_{min})
 \label{lab12}
 \end{equation}

 More specifically  the dynamics of $w (t)$ consists at each time $t$
in the updating  eq. (\ref{lab1})
 (or - equivalently - eq. (\ref{lab6})) {\bf except if} this results in
 $w(t+1)<w_{min}$ (or, equivalently $x (t+1) <  x_{min} = \ln (w_{min})$)
 in which case the updated new value is $w(t+1) = w_{min}$
 (respectively $x (t+1) = x_{min}$).

 These modifications are obviously not allowed by the Central Limit Theorem, and
consequently, the
 derivation in the previous section which led to the log-normal distribution
eqs. (13)(19) cannot be applied.

 Instead, one can obtain intuition on the modified system eq. (7)(20) by realizing
 that in fact the system eq. (\ref{lab6}) with the constraint
 eq. (\ref{lab12}) can be interpreted as the vertical motion of a molecule
in the earth gravitational field above the earth surface.
 In this case the ``earth surface'' is $x_{min}$, the
 gravitationally induced downward drift is $v = <x>/t = < \ln \lambda >$
 and the diffusion per unit time is parametrized by the squared standard deviation
 $\sigma^2_{\ln \lambda}$.

 For $v < 0$ this is the barometric problem  and it has the static solution:


 \begin{equation}
   P (x) \sim \exp { -x/kT}
   \label{lab13}
 \end{equation}
 which when re-expressed in terms of $w$'s becomes (cf. eq. (10)):
 \begin{equation}
   P (w) d w \sim   e^{- (\ln w)/kT} d \ln w
   \label{lab14}
 \end{equation}
 i.e.
 \begin{equation}
   P (w) \sim w^{-1 - 1/kT}
   \label{lab15}
 \end{equation}
In order to estimate  the value of the exponent
 \begin{equation}
   \alpha = 1/kT
   \label{lab16}
 \end{equation}
one can substitute the form eq. (\ref{lab15}) in the master equation
which governs the evolution of the  probability distribution of the process eq. (\ref{lab1}).
The master equation expresses the flow of probability between the
various values of $w$ as  $w(t)$ is updated to $w(t+1) = \lambda w(t) $.
More precisely, the probability for $w(t+1) =  \lambda w(t)$ to equal
a certain value  $w$ is the integral over $\lambda$
(weighted by $\Pi (\lambda)$)
of the probabilities  that $w(t) = w/\lambda$.
This leads in equilibrium to the relation:
 \begin{equation}
   P (w) = \int \Pi(\lambda) P (w/\lambda)  d(w/\lambda)
   \label{master}
 \end{equation}
i.e., substituting (24)(25) into eq. (26):
 \begin{equation}
w^{-1 - \alpha}  = \int \Pi(\lambda) (w/\lambda)^{-1 - \alpha}  d(w/\lambda)
   \label{master1}
 \end{equation}
or, by dividing both members by $w^{-1 - \alpha}$:
 \begin{equation}
1   = \int {\lambda}^{ \alpha} \Pi(\lambda) d\lambda
   \label{master2}
 \end{equation}
This means that the value of $\alpha$ is given by the transcendental equation:
(Solomon and Levy 1996):
 \begin{equation}
< {\lambda}^{ \alpha}> = 1
   \label{lab17}
 \end{equation}

One may wonder why this equation does not hold for the
problem in the previous section (the system eq. (11) without the lower bound eq. (20))
since the lower bound $w_{min}$ does not seem to appear in the formula
eq. (\ref{lab17}).
The answer is that in the absence of the lower bound, there is no
stationary equation (\ref{master}) as the system evolves forever
towards the non-normalizable  solution $P(w) \sim w^{-1} $.
In fact, formally, $\alpha = 0$ {\it is} a solution of eq. (\ref{lab17}).
This "solution"  is in fact the relevant one in the case  $< {\ln \lambda}> \ge 0$.

Another nontrivial fact is that the limit $w_{min} \rightarrow 0$
leads to $\alpha \rightarrow 1$ rather than
$\alpha \rightarrow 0$ (which is the value in the total absence
of a lower bound). This nonuniform behavior of the limits $N \rightarrow \infty$
and $w_{min} \rightarrow 0$ is related to the non-trivial thermodynamic
limit of the system (Biham et al. 1998).

One of the problems with the solution to eq. (\ref{lab17})
is that while it is independent on $w_{min}$,it is highly dependent
on the shape and position of the distribution $\Pi(\lambda)$
of the random factor  $\lambda (t)$ and consequently it is highly
dependent on the changes in the dynamics eq. (11).

In the systems introduced in the following sections, the situation will be
the opposite: the characteristics of $\Pi(\lambda)$ will be largely
irrelevant and the lower bound eq. (20) will play a central role.

In particular, the distribution of the social wealth $P (w)$ and the
inflation ($d {\bar w} / dt$) of the system will depend on
the social security policy,
i.e. on the poverty bound $w_{min}$ below of which the individuals are
subsidized (Anderson 1995).
The relevant parameter is the ratio
 \begin{equation}
   q  =  w_{min} / {\bar w}
   \label{lab18}
 \end{equation}
 between the minimal wealth $w_{min}$
and the average wealth:
 \begin{equation}
   \bar w = 1/N \sum_i w_i
   \label{lab19}
 \end{equation}
In the presence of inflation, it is $q$ rather then $ w_{min}$
which has to be fixed since
a fixed minimal wealth independent on the changes in the average wealth
would be not very effective for long time periods.

To prepare the formalism for those more sophisticated applications we
will deduce below an identity relating the exponent $\alpha = 1/kT$
to the ratio $q$ eq. (30).

 Such a relation can be obtained using in eq. (\ref{lab15})(25)
 the fact that the total probability is $1$:
 \begin{equation}
   \int_{w_{min}}^{\infty} P(w) dw = 1
   \label{lab20}
 \end{equation}
 i.e.
 \begin{equation}
   Const. \ \int_{w_{min}}^{\infty} w^{-1 -\alpha} dw = 1
   \label{lab21}
 \end{equation}
 and the fact that the average of $w$ is ${\bar w}$:
 \begin{equation}
   \int_{w_{min}}^{\infty} w P(w) dw = {\bar w}
   \label{lab22}
 \end{equation}
 i.e
 \begin{equation}
   Const. \ \int_{w_{min}}^{\infty} w^{-\alpha} dw = {\bar w}
   \label{lab23}
 \end{equation}
 By extracting $Const$ from eq. (\ref{lab21}):
 \begin{equation}
   Const = \alpha w_{min}^{\alpha}
   \label{lab24}
 \end{equation}
 and introducing this value  in eq. (\ref{lab23}) one gets the relation:
 \begin{equation}
   \alpha w_{min}^\alpha  [-w_{min}^{1 - \alpha}/(1 - \alpha) ] =  {\bar w}
   \label{lab25}
 \end{equation}
 By dividing both members by ${\bar w} {{\alpha}\over{1-\alpha}}$ one obtains:
 \begin{equation}
   -w_{min} / {\bar w} = 1/\alpha - 1
   \label{lab26}
 \end{equation}
 Considering the definition eq. (\ref{lab18}) this reads:
 \begin{equation}
   \alpha = 1 / (1 - q)
   \label{lab27}
 \end{equation}

This is a quite promising result because for the natural range
of the lower bound ratio $0 < q < 1/2$ it predicts $1 < \alpha < 2$
which is the range of exponents observed in nature (which is far from the
log-normal value $\alpha = 0$).

 The main problem with the power-low generating mechanism based
 on the single-agent dynamics eq. (11)(20) is that it works only for negative values
 of $<\ln \lambda>$
 (the barometric equation does not hold for
 a gravitational field directed upwards).

 This means that (except for the neighborhood of $w_{min}$) $w(t+1)$ is
 typically smaller than $w(t)$.  This is not the case in nature
 where populations, economies are expanding (at least for certain
 time periods).

 Moreover, the exponent $\alpha$ of the power law is highly unstable to
 fluctuations in the parameters of the system.
 In particular, trying to model large changes in
 $\bar w$ one is lead to large fluctuations of $q$ and consequently
 (cf. eq. (\ref{lab27}))
 to large variations in the exponent $\alpha$.  This again is in disagreement
 with the extreme stability of the exponents $\alpha$ observed in nature.

 We will see later how the multi-agent GLV solves these problems.
 The key feature is to introduce interactions between the individual
 elements $w_i$ through the inclusion of terms
(or lower bounds) proportional to  $\bar w$.

\section{Multiplicative processes coupled through
         the lower bound}

 In the previous section we showed that power-laws eq (\ref{lab15})
 can be obtained from multiplicative stochastic dynamics with lower
 bound in the same way that the exponential laws eq. (22)
 can be obtained in additive stochastic dynamics eq. (8)
 bounded from below eq. (21).

 The problematic points were: the instability of the exponent $\alpha$
 to variations in the average wealth/population and the fact that the mechanism
 accounts only for ``deflating''/``shrinking'' of $w$.

 However, the solution of these difficulties is already apparent in
 eq. (\ref{lab27}):  we consider a system of $N$
 degrees of freedom $w_i$
 (with $i=1,...N$) which is governed by the following dynamics:

 at each time $t$, one of the $w_i$'s is chosen randomly to be
 updated according to the formula:
 \begin{equation}
    w_i (t+1) = \lambda (t) w_i (t)
    \label{lab28}
 \end{equation}
 while all the other $w_i$'s are left unchanged.
 At each instance, the random factor $\lambda$ is extracted anew from
 an $i$-independent probability distribution $\Pi (\lambda)$.

 The only exception to the prescription eq. (\ref{lab28}) is if following
 the updating  eq. (\ref{lab28}) $w_i (t+1)$ (or other $w_k$'s)  end up
 less then a certain fixed fraction $0 < q < 1$ of the average $\bar w$:
 \begin{equation}
    w_i (t+1)  <  q {\bar w} (t)
    \label{lab29}
 \end{equation}
 The prescription, if eq. (\ref{lab29}) happens, is to further update
the affected $w_j$'s to:
 \begin{equation}
    w_j (t+1)  =  q {\bar w} (t) .
    \label{lab31}
 \end{equation}
 The wonderful property of the system eq. (\ref{lab28}--\ref{lab31})
 is that when re-expressed in terms of the variables
 \begin{equation}
    v_i (t) =  w_i (t) / {\bar w} (t)
    \label{lab32}
 \end{equation}
 it leads to a system very close to the system
 eq. (\ref{lab1}),(\ref{lab11}):
 \begin{equation}
    v_i (t+1) =  {\tilde \lambda} (t)  v_i (t)
    \label{lab33}
 \end{equation}
 \begin{equation}
    v_i (t+1) >  q
    \label{lab34}
 \end{equation}
 where the effective multiplicative factor in eq. (\ref{lab33}) is:
\begin{equation}
 {\tilde \lambda} (t) = \lambda (t) {\bar w(t)}/{\bar w(t+1)}
   \label{newlam}
 \end{equation}

Note that this solves the problem of $<\ln \lambda > > 0$  because
the multiplication with the (``renormalization'') factor
 ${\bar w}(t)/{\bar w}(t+1)$ takes care that the $v$ distribution is
never running to infinity (in fact it insures that ${\bar v} (t) = 1$ always).

 Another way to see how this solves the problems  with the single ``particle'' dynamics
 eqs. (\ref{lab1})(\ref{lab11})
is to realize that now, with a $q \bar w$ lower bound,
 even if the average ${\bar w} (t)$ runs to infinity, the lower bound runs after
 it in such a way as to insure a stationary value of $\alpha$.

 Indeed, according to eq. (\ref{lab15}), the $v_i$ dynamics
 eq. (\ref{lab33}--\ref{lab34}) leads to a $v_i$ distribution
 \begin{equation}
    P (v) \sim  v^{-1 -\alpha}
 \end{equation}
 which in turn implies
 \begin{equation}
    P (w) \sim  w^{-1 -\alpha}
 \end{equation}
Since we do not have a close analytic formula
for ${\tilde \lambda}$ eq. (46) in terms of the model parameters,
the transcendental equation  eq. (\ref{lab17}) is not useful here.

However, since $\bar v = 1 $ by definition,
one gets according eq. (\ref{lab27}) the following close
formula for the exponent $\alpha$:
 \begin{equation}
    \alpha = 1 / (1 - q)
 \end{equation}
 This means that even for significantly time-varying distributions
 ${\Pi}_t (\lambda)$ in eq. (40), the exponent of the power law
 remains time-invariant.

{\bf
 The above results are quite non-trivial in as far as they predict the
 stochastic behavior of highly interactive and time non-stationary
 systems eqs. (40)-(42) by relating them
 formally to non-interacting static statistical systems eqs. (7)(20).
}

We will continue this line in the following sections an reduce the
highly nontrivial GLV system to a rather simple
single-agent linear stochastic  equation.

\section{The single-agent linear stochastic equation}

 The discrete equation below is a more elaborate
version of the single-agent model introduced in Section 6.
Here the lower bound is
 supplied  effectively by an additive random term $\rho$, instead of
 an explicit barrier eq. (20):
 \begin{equation}
      w(t+1) = \lambda(t) w(t) + \rho(t)
\label{kesten}
 \end{equation}
 The equation (\ref{kesten}) may (crudely) describe the time-evolution of
 wealth for one trader in the stock market.  In this context
 $w$ may represent the wealth of one trader, who performs at each
 time-step a stock transaction.  $w(t)$ is the wealth at time $t$,
 and $w(t+1)$ is the wealth at time $t+1$.

 $\lambda$ and $\rho$ are random variables extracted from positive
 probability distributions.
 $\lambda$ may be called the ``success factor'', and $\rho$ is a
 ``restocking term'', which takes into account the wealth acquired
 from external sources (e.g. using the state built infra-structure,
 subsidies etc).

 The  dynamics eq. (\ref{kesten}) leads to a power-law distribution $P(w) \sim w^{-1-\alpha}$,
 if $<\ln \lambda > < 0$.
More precisely, the $P (w)$ has a power-tail for values
of $w$ for which $\rho(t)$ is negligible with respect to
$\lambda(t) w(t)$.
 If $<\ln \lambda> \ge 0$  the distribution is a log-normal expanding in time,
 which in the infinite time limit corresponds to a power-law
 with exponent $-1-\alpha = -1$.
I.e. the mechanism eq. (50) does not explain expanding economies with $\alpha > 0$.

 Since the distributions (and initial values of $w$) are positive,
 the $\rho(t)$ term keeps the value of $w(t)$ above a certain
 minimal value of order $\bar \rho$.  Therefore, for large
enough values of $w$ the dynamics is indistinguishable from our
 previous model, the multiplicative process (7)
 with fixed barrier eq. (\ref{lab11}).

It is therefore not surprising that it can be rigorously proven
(Kesten 1973) that eq. (\ref{kesten}) leads to a power law eq. (\ref{lab15})
with exponent $\alpha$ given by the transcendental equation (\ref{lab17}):
 \begin{equation}
< {\lambda}^{ \alpha}> = 1
   \label{labkest}
 \end{equation}
It is again notable that (as in the case of the $w_{min}$ lower bound),
 the exponent $\alpha$ is totally independent
on the distribution of the additive term $\rho$ while it is highly sensitive
to the shape and position of the $\Pi (\lambda )$ distribution.
We will see that in GLV the situation is reversed.

\section{ The Generalized Lotka-Volterra system}

 The GLV (Solomon and Levy 96) implements a more complex dynamics than the  eq. (\ref{kesten}).
 The advantages however overweight the difficulties:
 \begin{itemize}
   \item{}  The GLV model insures a stable exponent $\alpha$ of the power-law
$P(w) \sim w^{-1-\alpha}$ even
            in the presence of large fluctuations of the parameters.
   \item{}  The value of $<\lambda>$ and the average $\bar w$
 can vary during the run (and
            between the runs) without affecting the exponent $\alpha$ of the
            power-law distribution.
   \item{}  $\lambda$ can take typical values both larger or
            smaller than $1$.
 \end{itemize}

 \medskip
 The GLV is an interactive multi-agent model. We have $N$ traders, each having
 wealth $w_i$, and each $w_i$ is evolving in time according to:
 \begin{equation}
      {w_i(t+1)} = {\lambda(t) w_i(t) + a(t) {\bar w}(t)
- c(t) {\bar w}(t) w_i(t)}
\label{lastglv}
 \end{equation}
 Here $\bar w$ is the average wealth, which supplies the coupling
 between the traders:
 \begin{equation}
      {\bar w} = (w_1 + w_2 + \dots + w_N)/N
 \end{equation}
 $\lambda$ is a positive random variable with a probability distribution $\Pi (\lambda)$
 similar to the $\lambda$ success factor we used in eq. (50) in the previous section.
The dramatic difference is that now, $\lambda$ can take systematically
values larger than $1$ and in fact its distribution can vary in time
(and have time intervals with $< \ln \lambda >$ both smaller and larger than $0$).

 The coefficients $a$ and $c$ are in general functions of time,
 reflecting the changing conditions in the environment.

 The coefficient $a$ expresses the auto-catalytic property of wealth
 at the social level, i.e. it represents the wealth the individuals
 receive as members of the society in subsidies, services and social
 benefits.  That is the reason it is proportional to the average wealth.

 The coefficient $c$ originates in the competition between each
 individual and the rest of society.  It has the effect of limiting
 the growth of $\bar w$ to values sustainable for the current conditions
 and resources.

\section{Reducing GLV to a set of independent equations (50)}

 Summing the GLV eq. (\ref{lastglv}) over $i$,
and taking the local time average one gets
for the local time  average $<{\bar w}> (t)$
an equation  similar to the scalar LV equation eq. (1).
 \begin{equation}
    N <{\bar w}> = <\lambda> N <{\bar w}>  + <a(t)> N <{\bar w}> - <c(t)> N <{\bar w}>^{2}
 \end{equation}
which gives :
 \begin{equation}
   <{\bar w}> = {{ - 1 + <\lambda> +  <a(t)>} \over {<c(t)>}}
\label{aver}
 \end{equation}

Neglecting the fluctuations of the  average $<\bar w> (t) $
during the updating of a single individual $i$,
one can substitute $<\bar w> (t)$ for $\bar w$ it in the last term of
eq. (\ref{lastglv}) and regroup the terms linear in $w_i$:
 \begin{equation}
      {w_i(t+1)} =
      [ 1 + \lambda(t) - <\lambda(t)> -  <a(t)>] w_i(t) + a(t) {\bar w}(t)
\label{newlv}
 \end{equation}
 Introducing like we did in eq. (\ref{lab32})
 wealth values normalized by the average wealth:
 \begin{equation}
    v_i (t) =  w_i (t) / {\bar w}(t)
 \end{equation}
the equation (\ref{newlv}) becomes:
 \begin{equation}
      {v_i(t+1)} =
      {[1 + \lambda(t) - <\lambda(t)> -  <a(t)>)] v_i(t)} + a(t)
\label{newkest}
 \end{equation}
where we neglected again the fluctuation of $\bar w$ during the
time $t{\rightarrow}t+1$ and put ${ {{\bar w} (t) }\over {{\bar w} (t+1) }} = 1$.
 This can be justified rigorously
for $\alpha > 1$ in the large $N$ limit
because then the size of the largest $w_i$
is (cf. Solomon 1998) of order $ O({\bar w} N^{1-\alpha}) << 1$ and therefore the
changes in $\bar w (t)$ induced by any $w_i$ updating are negligible
of the leading order.
In finite systems $N < \infty$ and for $\alpha < 1$,
there are (computable) corrections.

The key observation now is that the system eq. (\ref{newkest})
has the form of $N$ decoupled equations of the form (\ref{kesten}) with
the effective multiplicative stochastic factor
 \begin{equation}
{\tilde \lambda} =
      1 +  \lambda(t) - <\lambda(t)> -  <a(t)>
\label{newkest1}
 \end{equation}
This means that
 the results of the single linear stochastic agent model eq. (\ref{kesten})
can be applied now to the $v_i(t)$'s in order to obtain
 \begin{equation}
    P (v) \sim  v^{-1 -\alpha}
 \end{equation}
 which in turn implies
 \begin{equation}
    P (w) \sim  w^{-1 -\alpha}
 \end{equation}
with the exponent $\alpha$ dictated by the equation eq. (\ref{labkest})(59):
\begin{equation}
    <  [1 + \lambda(t) - <\lambda(t)>  -  a(t)]^{\alpha} > =1
\label{newkest2}
 \end{equation}

This shows that the wealth distribution in a
economic model based on GLV is a power law.
In addition, it is easy to see using (58) that
the elementary steps in the time variations of the average wealth
are distributed by a (truncated) power law.
Accordingly, it was predicted that
 the market returns will be distributed by $L_{\alpha} ({\bar w})$
  a truncated Levy distribution
of index $\alpha$ (Solomon 1998). This turned out to be
 in accordance with the actual experimental data (Mantegna and Stanley 1996).

Note that similarly to the passage from the single agent system eq. (7)(20)
to the many agents system coupled by the lower bound eq. (40)(41),
here too, the {\bf formal} reduction of the GLV to a single agent system
implies very different properties in the actual "physical" system.

In particular while in the single agent  system the average of $\lambda$
was crucial in the fixing of the exponent $\alpha$ and in fact  $\lambda$ had to have
an average less then 1, in the multi agent model,
the average of $\lambda$ cancels in the expression (\ref{newkest1})
for $\tilde \lambda$. On the other hand, while the additive term
in eq (\ref{kesten}) had no role in
the fixing of $\alpha$, the corresponding term $a$ is one of the crucial factors
determining $\alpha$ in GLV (cf. eq. (\ref{newkest2})).
Moreover, while the parameters in the eq. (\ref{kesten}) model
allowed time variations in the $\bar w$ only at the price of variations in the
exponent $\alpha$, in the GLV system, one can arbitrarily change the
ecological/economic conditions $c$ so as to vary the total population/wealth
 by orders of magnitude (cf. eq. (55))
without affecting the power law and its exponent
($\bar \lambda$ is independent on $c$ and so is the solution of
$< {\tilde\lambda}^{\alpha} > = 1 $.

\section{The Financial interpretation of GLV}

In this section we discuss the various terms appearing in the equations,
their interpretation in the financial markets applications,
their effects and their implications for the financial markets phenomenology.

We first discuss the assumption that the individual investments/gains/losses
are proportional to the individual wealth:
$$ w(t+1) = \lambda w (t)$$
This is actually not true for the low income/wealth individuals
whom incomes do not originate in the stock market.
In fact the additive term $a \bar w$ tries to account for the
departures related to additional amounts originating in subsidies, salaries
and other fixed incomes.
However, for the range of wealth where one expects
power laws to hold ($w > \bar w$) it is well documented that the
investment policies, the investment decisions and the measured
yearly income are in fact proportional to the wealth itself.


The statistic uniformity of the relative gains and losses of
 the market participants is a weak form to express the
 fairness of the market and the lack arbitrage opportunities
(opportunities to obtain systematically higher gains $\lambda -1$
than the market average without  assuming higher risks):
for instance, if the distribution of $\lambda$ would be
systematically larger for small-$w$-investors,
then the large-$w$-investors would only have to split their wealth
in independently managed parts to mimic that low-$w$
superior performance. This would lead to an
equalization of their $\lambda$ to the  $\lambda$  of the low-$w$
investors. Therefore in the end, the distribution $\Pi (\lambda )$ will end up
$w$-independent as we assumed it to be in GLV from the beginning.

Almost every realistic microscopic market model we have
studied in the past
shares this characteristics of $w$-independent $\Pi (\lambda)$
distribution.

Turning to the terms relevant for the lower-bound $w$ region,
the assumption that
the average wealth contributes to the individual wealth
$a \bar w$ and the alternative mechanism assuming
a lower bound proportional to the average $q \bar w$
are both simplified mechanisms to prevent
the wealth to decrease indefinitely.

In practice they might be objectionable:
it is not very clear that the state subsidies can be invoked
to save bankrupt investors (though this happens often when large important
employers are at risk or when their collapse would endanger the stability
of the entire system).
In any case the term $a \bar w$ is appropriately taking into account
the arbitrariness of the money denominations.
More precisely, assume that (by inflation or by currency renaming)
the nominal value of all the money in the economy becomes 10 times
larger. Then the subsidies term  $a \bar w$ will become 10 times larger
preserving in this way the actual absolute value.
So to speak, multiplying a quantity by  $\bar w$ expresses it in
"absolute currency".

These mechanisms controlling the lower bound behavior
of the system may be considered as just parametrizations of
the continuum flow of investors/capital
to and from the large-$w$ investor ranges
relevant for the financial markets trading.
It is however possible that the subsidies to the very poorest
tumble their way (through the multiplicative random walk)
into the middle classes and end-up in the large-$w$ scaling region
in the way our models suggest.

More theoretical and experimental research is necessary in order
to discriminate between the various alternatives (or in order
to recognize them as facets of the same phenomenon).

For instance  one can  look at
the relation between $q$ and $\alpha$ as basically kinematic
in the sense that given the power law, it is unavoidable
that the lower bound $q \bar w$ would govern the exponent of the power law.

It would still be  interesting to study in detail how
the wealth pumped at the lower-bound barrier makes its way
to the large $w$ tail and or, alternatively in the case of
stationary $\bar w$ to find the way in which the additive
subsidies to the low-$w$ individuals in our models are covered by the
"middle classes".

For instance in the $q \bar w$ lower-bound model,
the dominant mechanism of extracting wealth from the
middle class seems to be the accelerated inflation:
the updating of the individual $w_i$'s
induces changes in $\bar w$ which in turn induces
changes in the position of the lower bound $q \bar w$ which leads to the
necessity to subsidize immediately all the $w_i$'s
situated between the old and the new poverty line.
This in turn leads to increase in the $\bar w$ and to the
completion of the positive feedback loop.
When looking at the normalized wealth $v_i = w_i / {\bar w}$,
this accelerated inflation is effectively a proportional tax
which reduces the $\lambda$ gains to smaller relative gains
$\tilde \lambda = \lambda w(t)/w(t+1)$.

In the case of the GLV model, there is no inflation (for constant $c$,
$\bar w $ is - modulo local fluctuations - constant in time).
The extraction from the middle class of the money for subsidies
is quite explicit from the way the subsidies term $a \bar w$ is affecting
negatively ($-a$) the $\tilde \lambda $ gain value eq. (39).
However most of the loss in $\tilde \lambda$ eq. (39) is
the term ($-<\lambda>$) and is due mainly to the
competition between the large traders $- c w_i \bar w$.
This term is in fact equivalent with enforcing a proportional tax.
The best way to see it is to recall from above analysis that multiplying the
wealth-proportional quantity $c w_i$ by $\bar w$ expresses it
in "absolute currency".

It would appear that the government has the choice of enforcing
such a proportional tax and keep $\bar w$ roughly constant,
or just print the money which it dispenses to the poor and
let the inflation tax the other classes.
One can of course make compromises between these 2 extremes by
allowing both taxation and inflation.
In any case, the net result can be only variations
in the relative size of
the middle class (variations of $\alpha$)
as the generic emergence of the
power law will be very difficult to avoid.

A government may try to get a more equalitarian distributions
(increasing $\alpha$) by increasing  $q$.
However this would bring an even larger population
in the neighborhood of the poverty line.
This would require more and more frequent subsidies to enforce
and consequently (according to the analysis above)
larger taxes and/or faster inflation.

Fearing this, another government might decide for low values of $\alpha$.
This would not only mean an increase of
the ratio between the richest and poorest
which might be morally questionable
but also lead to dramatically unstable fluctuations
in the market (e.g. in $\bar w$).
Indeed, one can show that typically
the largest trader owns $O (N^{1-\alpha})$ of the total wealth.
For finite $N$ and very low $q$, $\alpha$ may be shown to drop
below $1$ (contrary to eq. (49)). This would imply that almost all the
wealth is owned by the largest $w_i$.
It is well know however that the single agent discrete logistic map eq.
(1) leads generically to chaotic unstable dynamical regime (May 1976).

In between these extremes it might be that there is no much choice
of dynamically consistent $\alpha$ values outside the
experimentally measured range $1.4 < \alpha < 1.7$.
These values correspond (through eq. (49))
to poverty lines between $ 0.3 < q < 0.4$.
Larger values of $q$ would imply that almost everybody is subsidized
while a significantly lower value would mean the poorest
cannot literally live:
the average wages in an economy are automatically tuned as
to insure that a family of 3 can fulfill the needs considered basic
by the society on a one-salary income.
Somebody earning less than ${{1}\over{3}} \bar w$
will therefore have serious
problems to live a normal life.
In fact, such a person might be exposed to hunger
(even in very rich economies)
since even the prices of the basic food
are tuned to the level of affordability of the average family
in the given economy.

Far from implying a fatalistic attitude of the economic facts of life,
the analysis of our simplified models might lead
in more realistic instances to ideas and prescriptions
for dynamically steering the social economic policies
to optimal parameters both from the technical/efficiency
and from the human/moral point of view.

They might help transcend (through unification)
the traditional dichotomy by which
science is only in charge with deciding true from false while
humanities are in charge only with discerning good from bad.

\section{Further Economics applications of GLV}

As mentioned a few times, the GLV systems can be applied to
many power laws.

Even within the financial framework there are a few
apparently different realizations
of the GLV dynamics and of the power laws.

 For instance, one can consider the market as a
 set of companies $i = 1, ..... , N$
 whose shares are traded and whose prices vary in time
 accordingly.

 One can interpret then $w_i$
as the capitalization of the company $i$ i.e.
the total wealth of all the market shares of the company.

The time evolution of $w_i$ can still be represented by eq. (\ref{GLV}).
In this case, $\lambda$ represents the fluctuations
 in the market worth of the company. For a given total number of shares
 $\lambda$ is measured by the change in the individual share price.
   These changes take place during individual transactions
   and are typically fractions of the nominal share price
   (measured in percents or in points).

 With such an interpretation, $a$ represents the
 correlation between the worth of each company and the market index.
 This correlation is similar for entire classes of shares
 and differences in the $a$'s of various economic sectors
 result in the modern portfolio theory in "risk premia" which affect
 their effective returns $\tilde \lambda$ in a way similar to our formulae
 in a previous section
 (low correlation $a$ corresponds to larger effective incomes).

 $c$ represents the competition between the companies for the finite
 amount of money in the market
 (and express also the limits in their own absolute worth).
 We do not need to consider $c$ a constant.
 Time increases in the resources may lead to lower values of $c$ which in turn
 lead to increases in $\bar w$. However, as seen before, such changes
 do not affect the exponent of the power law distribution.

 In this interpretation, the GLV model will predict
 the emergence of a power-law in
 the probability distribution of company sizes (capitalization).
 In particular, this would imply that the weights of the various
 companies composing the S$\&$P 500 are distributed by a power law.
 In turn this would imply that the S$\&$P fluctuations follow a
 truncated Levy distribution of corresponding index.

 Yet another interpretation is to consider $w_i$ as the size of
 coordinated trader sets
 (i.e. the number of traders adopting a similar investment policy)
 and
 assuming that the sizes of these sets vary self-catalytically
 according to the random factor $\lambda$ while the $a$ term
 represents the diffusion of traders between the sets.
 Such an autocatalytic dynamics for the trading schools
 is not surprising as their decision processes involve elements similar
 to the use of common language, common values, which are
 central in the dynamics of languages and nations (which fulfill
 power laws) too.   The nonlinear
 term $c$ represents then the competition between these investing
 schools for individual traders membership.
 In fact if $a$ and $c$ fulfill the relation
 $a/c = \bar w$, the corresponding two terms in GLV taken together represent
 (in "absolute currency" $\bar w$) the act of each of the schools
 loosing (shedding) a certain $c$ fraction of its followers which are then
 spread uniformly between the schools.

 Such an interpretation of the GLV
 would predict a power-law in the distribution
 of trader schools sizes (similar to countries sizes).
 Again, such a distribution would account for
 the truncated Levy distribution of market fluctuations (with index
 equal to the exponent of the power-law governing trader sets sizes).

 It would be interesting to discriminate between the validity of
 the various GLV interpretations by studying experimentally the market
 fluctuations and the distributions of respectively individual wealth,
 companies capitalization and correlated investors sets.
 It is not ruled out that some of these interpretations may be
 consistent one with the other. This would be consistent in turn with the
modern portfolio theory claim on the existence of a "market portfolio"
 (stochastically) common to most of the traders.

 Further developments of the GLV model will include introducing variable
 number of agents $N$, studying the role of the discretization of the
 $w_i$ changes (due to indivisible units like people, shares, etc),
 taking into account the influence of the history of $w (t)$ on
 $\lambda$ (like in the case of investing strategies, crashes memory, etc.).

\section{Acknowledgements}

The research was supported in part by the Germany-Israel Foundation.
My thinking on the subject has been influenced by many of the
colleagues which share similar concepts (see refs.):
P. W. Anderson, A. Agay, P. Bak, J-P. Bouchaud, R. Cont, F. Fucito,
K. Illinsky, H. Levy, M. Levy, G. Mack, R. Mantegna, M. Marsili, S. Maslov,
R. Savit, D. Sornette,  H.E. Stanley, D. Stauffer, A. Weigend, G. Weisbuch,
Y-C. Zhang.

I benefited also from many illuminating discussions
on the subject with Prof. Shlomo Alexander
who will be always remembered as a brilliant scientist,
a broad intellectual and a symbol of the eternal youth.


\begin{thebibliography}{15}

 \bibitem[]{Anderson88}
 P. W. Anderson, J. Arrow and D. Pines, eds. The Economy as an Evolving
 Complex System (Redwood City, Calif.: Addison-Wesley, 1988);

 \bibitem[]{Anderson95}
P. W. Anderson in The Economy as an Evolving Complex System II
(Redwood City, Calif.: Addison-Wesley, 1995),
   eds. W. Brian Arthur, Steven N. Durlauf, and David A. Lane

 \bibitem[]{Atkinson78}
 A.B. Atkinson and A.J. Harrison, Distribution of Total Wealth in Britain
 (Cambridge University Press, Cambridge, 1978).

 \bibitem[]{Biham98}
 O. Biham, O. Malcai, M. Levy, S. Solomon, Phys. Rev. E \textbf{58}, 1352 (1998)

\bibitem[]{ Baillie and  Bollerslev90}
R.Baillie and T.Bollerslev,
Rev. of Econ. Studies 58 (1990) 565

\bibitem[]{ Bak 97}
P. Bak  (1997), How Nature Works, Oxford University Press, New York.

 \bibitem[]{Cahalan89}
 R. Cahalan and J. Joseph, Monthly Weather Review \textbf{117}, 261 (1989)

 \bibitem[]{Champernowne53}
 D. G. Champernowne, Econometrica \textbf{63}, 318 (1953)


 \bibitem[]{Ijiri77}
 Y. Ijiri and H. A. Simon,
 {\it Skew Distributions and the Sizes of Business Firms}
 (North-Holland, Amsterdam, 1977)

 \bibitem[]{Kesten73}
{H. Kesten, Acta Math. \textbf{131}, 207 (1973)}

\bibitem[]{Mark73}
KIM G.R. \& MARKOWITZ H.M. (1989) J. Portfolio Management, Fall 1989, 45.

\bibitem[]{May76}
{R.M. May, Nature 261 (1976) 207}

 \bibitem[]{Mantegna94}
 R. N. Mantegna and H. E. Stanley, Phys. Rev. Lett. \textbf{73}, 2946 (1994)

 \bibitem[]{Mantegna95}
 R. N. Mantegna and H. E. Stanley, Nature \textbf{376}, 46 (1995)

 \bibitem[]{Mantegna96}
 R. N. Mantegna and H. E. Stanley, Nature \textbf{383}, 587 (1996)

 \bibitem[]{Mantegna97}
 R. N. Mantegna and H. E. Stanley, Physica A \textbf{239}, 255 (1997)

 \bibitem[]{Mandelbrot51}
 B. Mandelbrot, Comptes Rendus \textbf{232}, 1638 (1951)

 \bibitem[]{Mandelbrot61}
 B. Mandelbrot, Econometrica \textbf{29}, 517 (1961)

 \bibitem[]{Mandelbrot63}
 B. Mandelbrot, J. Business \textbf{36}, 394 (1963)

 \bibitem[]{Pareto97}
 V. Pareto, Cours d'Economique Politique, Vol 2 (1897)

 \bibitem[]{Redner90}
 S. Redner, Am. J. Phys. 58 (3)

 \bibitem[]{Shlesinger82}
 M. F. Shlesinger and E. W. Montroll,
 Proc. Nat. Acad. Sci. USA (Appl. Math. Sci.)
 \textbf{79}, 3380 (1982)

 \bibitem[]{Simon58}
 H. A. Simon and C. P. Bonini, Amer. Econ. Rev. \textbf{48}, 607 (1958)

 \bibitem[]{Solomon95}
 S. Solomon, Ann. Rev. Comp. Phys. II p243 (World Scientific 1995) ed. D. Stauffer

 \bibitem[]{Solomon96}
 S. Solomon and M. Levy, J. Mod. Phys. C \textbf{7}, 745 (1996);
adap-org/9609002

 \bibitem[]{Solomon98}
 S. Solomon, Computational Finance 97 (Kluwer Academic Publishers 1998)
eds. A.P.-N. Refens A.N. Burgess and J.E. Moody

 \bibitem[]{Stanley95}
 M. H. R. Stanley, S. V. Buldyrev, S. Havlin, R. Mantegna, M. A. Salinger
 and H. E. Stanley, Europhys. Lett. \textbf{49}, 453 (1995)

 \bibitem[]{t'Hooft79}
G. t'Hooft, The proceedings of Cargese Summer Inst (1979) 135

 \bibitem[]{Tsallis}
 C. Tsallis, J. Stat. Phys. \textbf{52}, 479 (1988).

 \bibitem[]{Yule24}
 U. G. Yule, Phil. Trans. B. \textbf{213}, 21 (1924)

 \bibitem[]{Zanette97}
 D. H. Zanette and S. C. Manrubia, Phys. Rev. Lett. \textbf{79}, 523 (1997)

\bibitem[]{ZhangWB91}
Zhang W.B. (1991) {\it Synergetic Economics}. Springer, Berlin-Heidelberg.

\bibitem[]{Hellthaler}
T. Hellthaler Int. J. Mod. Phys. C, vol 6, (1995) 845
\\
Sex, Money, War and Computers:  Nontraditional Applications of
Computational Statistical Mechanics; S. Moss De Oliveira,  P.M.C
de Oliveira, D. Stauffer;  Springer-Verlag 1998
\\
M. H. R. Stanley, L. A. N. Amaral, S. V. Buldyrev, S. Havlin, H. Leschhorn, P.
Maass, M.  A. Salinger, and H. E Stanley; Nature 1996, vol 379, p 804

\bibitem[]{Zipf49}
 G. K. Zipf, Human Behavior and the Principle of Least Effort
              (Addison-Wesley Press, Cambridge, MA, 1949)

\bibitem[]{Zipf49}
     T. Lux "The Socio-Economic Dynamics of Speculative Markets: Interacting Agents, Chaos,
      and the Fat Tails of Return Distributions", in: Journal of Economic Behavior and
      Organization Vol. 33 (1998), pp. 143 - 165
 \\     "Time Variation of Second Moments from a Noise Trader/Infection Model", in:
  \\     Journal of Economic Dynamics and Control Vol. 22 (1997), pp. 1 - 3
   \\  "The Stable Paretian Hypothesis and the Frequency of Large Returns: An Examination
      of Major German Stocks", in: Applied Financial Economics Vol. 6 (1996), pp. 463
     \\  "Long-term Stochastic Dependence in Financial Prices: Evidence from the German
       Stock Market", in: Applied Economics Letters Vol. 3 (1996), pp. 701 - 706
\bibitem[]{Zipf49}
  D. Sornette and A. Johansen, Physics A, vol 245 (1997) 411
  \\   D. Sornette: Multiplicative processes and power laws (1998) Physical Review E
     57, p 4811-4813
    \\ D. Stauffer and D. Sornette: Log-periodic oscillations for biased diffusion on
     random lattice Physica A 252 (1998) 271-277
\bibitem[]{Zipf49}
     Takayasu H., Miura H., Hirabayashi T. and Hamada K.: Statistical properties of
     deterministic threshold elements - the case of market price, Physica A 184
     (1992) 127
   \\   Takayasu H., Sato A.-H. and Takayasu M.: Stable infinite fluctuations in
      randomly amplified Langevin systems, Phys. Rev. Lett. 79 (1997) 966
     \\ Takayasu H. and Okuyama K.: Country dependence on company size
     distributions and a numerical model based on competition and cooperation,
     Fractals 6 (1998) 67
    \\  Fractals in the Physical Sciences H. Takayasu J. Wiley \& Sons Chichester 1990
\bibitem[]{Zipf49}
      N.Vandewalle and M.Ausloos: Coherent and random sequences in financial
     fluctuations, Physica A 246 (1997) 454
    \\  N. Vandewalle, M Ausloos: Multi-affine analysis of typical currency exchange
     rates, The European Physical Journal B 4 (1998) 257-261
  \\    N. Vandewalle, M. Ausloos, Ph. Boveroux and A. Minguet: How the financial
     crash of october 1997 could have been predicted, The European Physical Journal
     B 4 (1998) 139-143
  \\   N. Vandewalle, P. Boveroux, A. Minguet, M. Ausloos: The crash of October 1987
     seen as a phase transition: amplitude and universality, Physica A 255 (1998) 201
  \\   N.Vandewalle and M.Ausloos: Sparseness and Roughness of foreign exchange
     rates, Int. J. Mod. Phys. C 9 (1998) 711
   \\ N.Vandewalle and M.Ausloos: Extended Detrended Fluctuation Analysis for
     financial data,Int. J. Comput. Anticipat. Syst. 1 (1998) 342
\bibitem[]{Zipf49}
      M. Marsili, S. Masvlov and Y.-C. Zhang: Dynamical optimization thery of a
     divesified portfolio, Physica A 253 (1998) 403-418
    \\ S. Galluccio, G. Caldarelli, M. Marsili, Y.-C. Zhang: Scaling in currency exchange,
     Physica A 245 (1997) 423-436
   \\  G. Caldarelli, G. Marsili, Y.-C. Zhang: A Prototype Model of Stock
     Exchange, cond-mat/9709118, Europhysics Letters 40, p479

\bibitem[]{Sornette97a}
{D. Sornette and R. Cont }, {J. Phys. I France 7 (1997) 431}

\bibitem[]{Bak97}
{P. Bak, M. Paczuski and M. Shubik}, {Physica A 246, Dec. 1, No 3-4, 430 (1997).
}
                                      

\end{thebibliography}
\end{document}